\documentclass[12pt]{article}
\usepackage[super,compress]{cite}
\usepackage{amsmath}
\usepackage{amssymb}
\usepackage{graphicx}
\begin{document}

\begin{center}
{\bf
Differential elastic nucleus-nucleus scattering in complete
Glauber theory}

Yu.M. Shabelski and A.G. Shuvaev 

\vspace{.5cm}

Petersburg Nuclear Physics Institute, Kurchatov National
Research Center\\
Gatchina, St. Petersburg 188300, Russia\\
\vskip 0.9 truecm
E-mail: shabelsk@thd.pnpi.spb.ru\\
E-mail: shuvaev@thd.pnpi.spb.ru

\end{center}

\vspace{1.2cm}

\begin{abstract}
\noindent
The differential elastic cross sections of $^{12}$C -- $^{12}$C
and $^{11}$Li -- $^{12}$C
nuclei are calculated in the complete Glauber theory.
The role of the possible correlations connected to the shell effects
in $^{11}$Li nucleus is considered.
\end{abstract}

\section{Introduction}
High energy nuclear interactions are usually treated in
the Glauber theory~\cite{Glaub1,Glauber:1970jm, Czyz:1969jg}.
It is highly efficient in describing
the hadron-nucleus scattering where all the relevant diagrams
can be summed up in the closed form. The nucleus-nucleus scattering
becomes rapidly more complicated for $A>4$ and additional simplified
approximations are usually made to get the result analytically.
An example of such calculations for the differential elastic
and reaction cross sections for $^{11}$Li -- $^{12}$C scattering
can be found in Refs.~\cite{Alkhazov:2011ks, Alkhazov:2013mqa}.

In this paper the analytical calculations of the differential
elastic cross sections for $^{12}$C -- $^{12}$C and $^{11}$Li -- $^{12}$C
scattering have been carried out in the complete Glauber theory
in the generating function approach Ref.~\cite{Shabelski:2021iqk}.
The results have been obtained for several commonly used forms of
the nuclear density, which allows one to get an insight to
the possible nuclear correlations, in particular, to those
related to the shell structure of the halo nucleus $^{11}$Li.

It is worth pointing out that it is the differential cross section
rather than the integrated reaction one that is more sensitive
to the nuclear density distribution, especially to its far distance parts.

\section{Differential elastic cross sections}

The amplitude of the elastic scattering of the incident nucleus
$A$ on the fixed target nucleus $B$ is given in the Glauber
theory by the expression~\cite{Pajares:1983gw, Braun:1988pk}
\begin{equation}
\label{dcs}
F_{AB}^{el}(q)\,=\,\frac{ik}{2\pi}\int d^{\,2}b\,e^{iqb}
\bigl[1\,-\,S_{AB}(b)\bigr],
\end{equation}
where $q$ is the transferred momentum and $k$ is the mean
nucleon momentum in nucleus $A$. The two-dimensional impact
momentum $b$ lies in the transverse plain to the vector $k$.
The function $S_{AB}(b)$ is evaluated through the sum
of the independent phase shifts $\chi_{NN}(b)$ for each
nucleon-nucleon scattering,
\begin{equation}
\label{I}
S_{AB}(b)\,=\,\langle\,A,\,B\,|
\left\{\prod\limits_{i\,j}\bigl[1-\Gamma_{NN}(b+x_i-y_j)\bigr]
\right\}
|\,A,\,B\,\rangle,
\end{equation}
with
\begin{equation}
\label{chi}
\Gamma_{NN}(b)\,=\,1\,-\,e^{i\chi_{NN}(b)}\,=\,
\frac 1{2\pi ik}\int d^{\,2}q\,e^{-iqb}f_{NN}^{el}(q).
\end{equation}
The symbol $\langle\,A,\,B\,|\,\cdots\, |\,A,\,B\,\rangle$
denotes an average over the nucleons' positions
$x_i$ and $y_j$ in the transverse plain. Each pair
$\{i,j\}$ enters the product only once, meaning that
each nucleon from the projectile nucleus can scatter
on each nucleon no more than once.

The elastic nucleon-nucleon amplitude, $f_{NN}$, is mainly
imaginary at the beam energy about 1~GeV per nucleon,
$\mathrm{Re} f_{NN}/\mathrm{Im} f_{NN}\lesssim 10^{-1}$.
The standard parametrization is
\begin{equation}
\label{f_NN}
f_{NN}^{el}(q)\,=\,ik\frac{\sigma_{NN}^{tot}}{4\pi}
e^{-\frac 12\beta q^2},
\end{equation}
where $\sigma_{NN}^{tot}$ is the total nucleon-nucleon
cross section. The slope $\beta$ is related to an effective
interaction radius $a = \sqrt{2\pi\beta}$.
\begin{figure}[htb]
\includegraphics[width=.9\hsize]{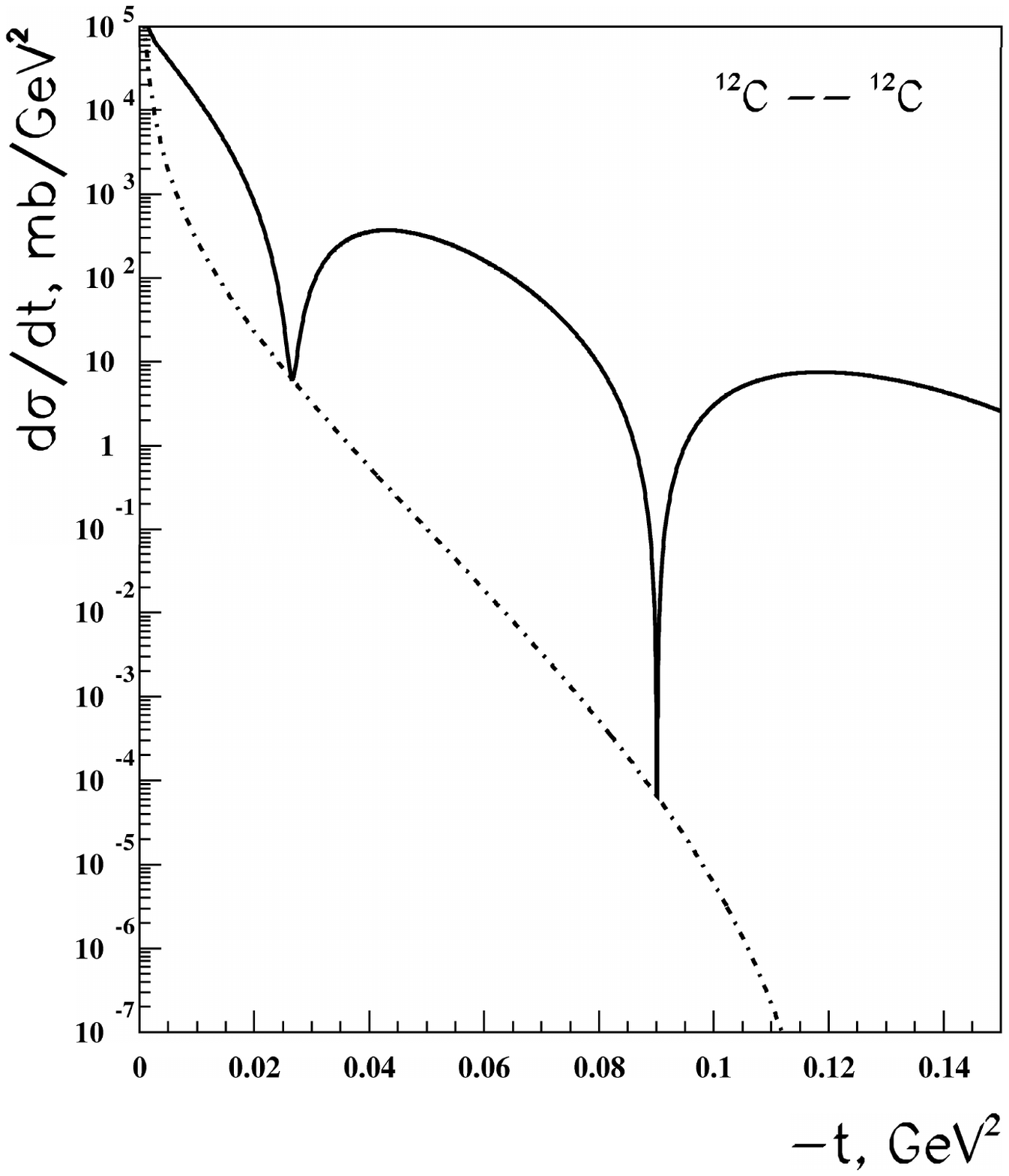}
\vskip -2.cm \caption{\footnotesize Differential elastic cross section of $^{12}$C-
$^{12}$C scattering at the energy about 1~GeV per nucleon. The dash-dotted line
shows the pure Coulomb scattering,
the solid line is for the sum of the Coulomb
and strong scatterings, Eq.(\ref{sum}).
}
\end{figure}
An efficient way to deal with the function (\ref{I})
is provided by the generating function $Z(u,v)$,
\begin{equation}
\label{ddZ}
S_{AB}(b)\,=\,\frac 1{Z(0,0)}
\frac{\partial^A}{\partial u^A}
\frac{\partial^B}{\partial v^B}\,
Z(u,v)\biggl|_{u=v=0},
\end{equation}
for which the closed expression have been obtained
in Ref.~\cite{Shabelski:2021iqk}
\begin{eqnarray}
\label{Zuv}
Z(u,v)\,&=&\,e^{W_y(u,v)},~~~~~~~~
z_y\,=\,1-\frac 12 \frac{\sigma_{NN}^{tot}}{a^2},
\\
\label{Wy}
W_y(u,v)\,&=&\, \frac 1{a^2}\int d^{\,2}x\,
\ln\bigl(\!\!
\sum\limits_{M\le A,N\le B}
\frac{z_y^{M\, N}}{M!N!}
\bigl[a^2 u\rho_A^\bot(x-b)\bigr]^M
\bigl[a^2 v\rho_B^\bot(x)\bigr]^N
\bigr).
\end{eqnarray}
The transverse densities entering this formula are expressed
through the three dimensional nucleon distributions in
the colliding nuclei,
$$
\rho_{A,B}^\bot(x_\perp)\,=\,
\int d z\,\rho_{A,B}(z,x_\perp),~~~
\int d^2 x_\perp\,\rho_{A,B}^\bot(x_\perp)\,=\,1.
$$
The function $W_y(u,v)$ (\ref{ddZ})
goes as the series built of the overlaps,
\begin{equation}
\label{tmn}
t_{m,n}(b)=
\frac 1{a^2}\int d^{\,2}x\,
\bigl[a^2\rho_A^\bot(x-b)\bigr]^m\,
\bigl[a^2\rho_B^\bot(x)\bigr]^n,
\end{equation}
with $m\le A$ and $n\le B$
For relatively light nuclei, say, $A,B\lesssim 20$,
the generating function can be explicitly differentiating
with respect $u$, $v$ variables yielding the final
$S_{AB}(b)$ function. For more heavy nuclei the differentiating
amounts with a rather good accuracy to sending these variables to
$A$ and $B$ values
(see Ref.~\cite{Shabelski:2021iqk} for details).

The density distribution
in a relatively light nuclei with the atomic weight
$A\lesssim 20$ is
well parametrized by the harmonic oscillator
function
\begin{equation}
\label{HO}
\rho_A(r)\,=\,\rho_0\bigl[1+\frac 16\,(A-4)\frac{r^2}{\lambda^2}
\bigr]e^{-\frac{r^2}{\lambda^2}},
\end{equation}
with $\rho_0$ being the normalization, and the factor $\lambda$
being adjusted to match the nuclear mean square radius,
$R_{\mathrm{rms}}=\sqrt{r_A^2}$,
$r_A^2=\int d^3r\,r^2\rho_A(r)$. The simple form
of this parametrization enables one to get analytically
all the overlaps relevant to $^{12}$C --$^{12}$C
scattering.
For more complicated densities the overlaps
can be calculated numerically.

There exist several parametrizations for $^{11}$Li nucleon density, which are more
complex due to the halo structure. The density is supposed to be the sum of two
separate parts for the central core and for the surrounding it neutron
halo~\cite{Alkhazov:2004syy, Ilieva:2012vhs, Korsheninnikov:1997qta,
Zhukov:1993aw}. Below we use the particular parametrization of the
form~\cite{Hassan:2015dfa}
\begin{equation}
\label{halo}
\rho(r)\,=\,N_c\rho_c(r)\,+\,N_v\rho_v(r),
\end{equation}
$N_c$ is the number of the nucleons
in the core, and $N_v$ is the number of the valence neutrons
in the halo. The core density is taken
in Gaussian form,
$$
\rho_c(r)\,=\,\frac 1{\pi^{\frac 32}a_c^3}\,e^{-\frac{r^2}{a_c^2}},
~~~~a_c=\sqrt{2/3}\,R_c,
$$
where $R_c$ is the core mean square radius. The halo is also parametrized
in the Gaussian forms, which are different for the various shell structures
supposed for it,
\begin{equation}
\label{params}
\begin{array}{ccc}
  \rho_v^G(r)\,=\,\frac 1{\pi^{\frac 32}a_G^3}e^{-\frac{r^2}{a_G^2}},~~~&
  ~\rho_v^O(r)\,=\,\frac 2{3\pi^{\frac 32}a_O^5}\,r^2e^{-\frac{r^2}{a_O^2}},~~~&
  ~\rho_v^{2S}(r)\,=\,\frac 2{3\pi^{\frac 32}a_{2S}^3}
   \left(\frac{r^2}{a_{2S}^2}-\frac 32\right)^2
  e^{-\frac{r^2}{a_{2S}^2}},
 \\
 a_G=\sqrt{2/3}\,R_v,  & a_O=\sqrt{2/5}\,R_v, & a_{2S}=\sqrt{2/7}R_v,
\end{array}
\end{equation}
where $R_v$ is the halo mean square radius. Whilst more
complicated, these forms also admit the straightforward
analytical evaluation of $t_{m,n}(b)$ functions.
\begin{figure}[htb]
\includegraphics[width=.9\hsize]{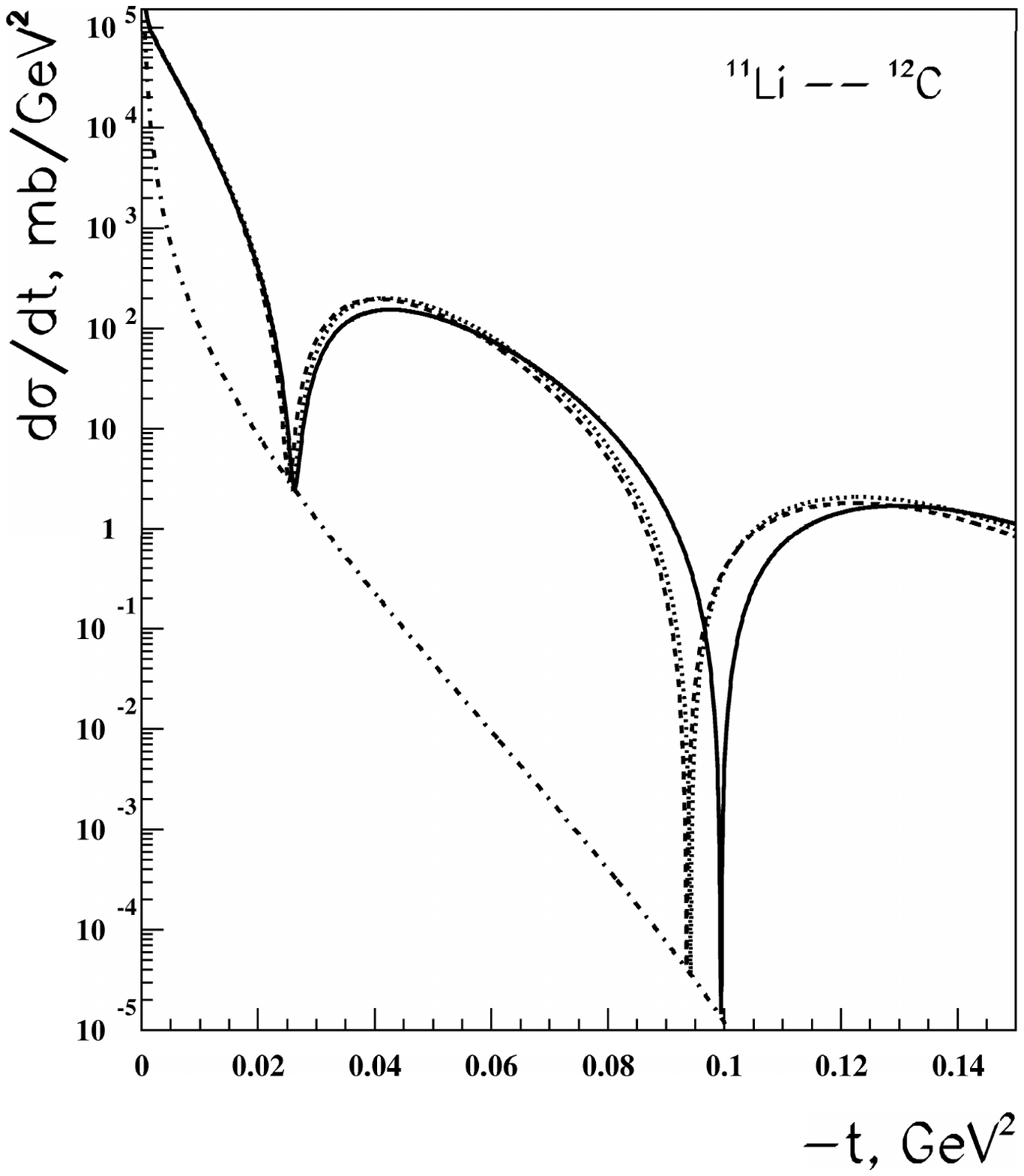}
\vskip -2.cm
\caption{\footnotesize
Differential elastic cross section of $^{11}$Li--$^{12}$C
scattering at the energy about 1~GeV per nucleon.
The different lines are for the sum of the strong and Coulomb
cross sections for the various halo
density parametrizations~(\ref{params}). The dotted line is for
$\rho_v^G$, the dashed line is for $\rho_v^O$, the solid line is for
$\rho_v^{2S}$.
The dash-dotted line stands for the pure Coulomb cross section.
}
\end{figure}

It is also necessary to account for the Coulomb scattering.
The Coulomb amplitude reads
$$
f_C(q)\,=\,-\frac 1{2\pi}M_{AB}\,Z_A Z_B e^2
\int d^3z\,e^{iqz}V_c(z),~~~
V_c(z)\,=\,\int d^3x d^3y\,\frac{\rho_A(x)\rho_B(y-z)}{|x-y|}
$$
where $q=p-p^\prime$ is the momentum transfer, $M_{AB}$
is reduced mass of the A and B nuclei, $Z_Z$, $Z_B$ are
their charge. Passing to the form factors one gets
\begin{equation}
\label{Coulomb}
f_C(q)\,=\,-2M_{AB}Z_A Z_B e^2\,\frac{\rho_A(q)\rho_B(q)}{q^2},
~~~\rho_{A,B}(q)\,=\,\int d^3x\,e^{iqx}\rho_{A,B}(x),
\end{equation}
the core form factor having been taken for the halo nucleus.
\begin{figure}[htb]
\includegraphics[width=.8\hsize]{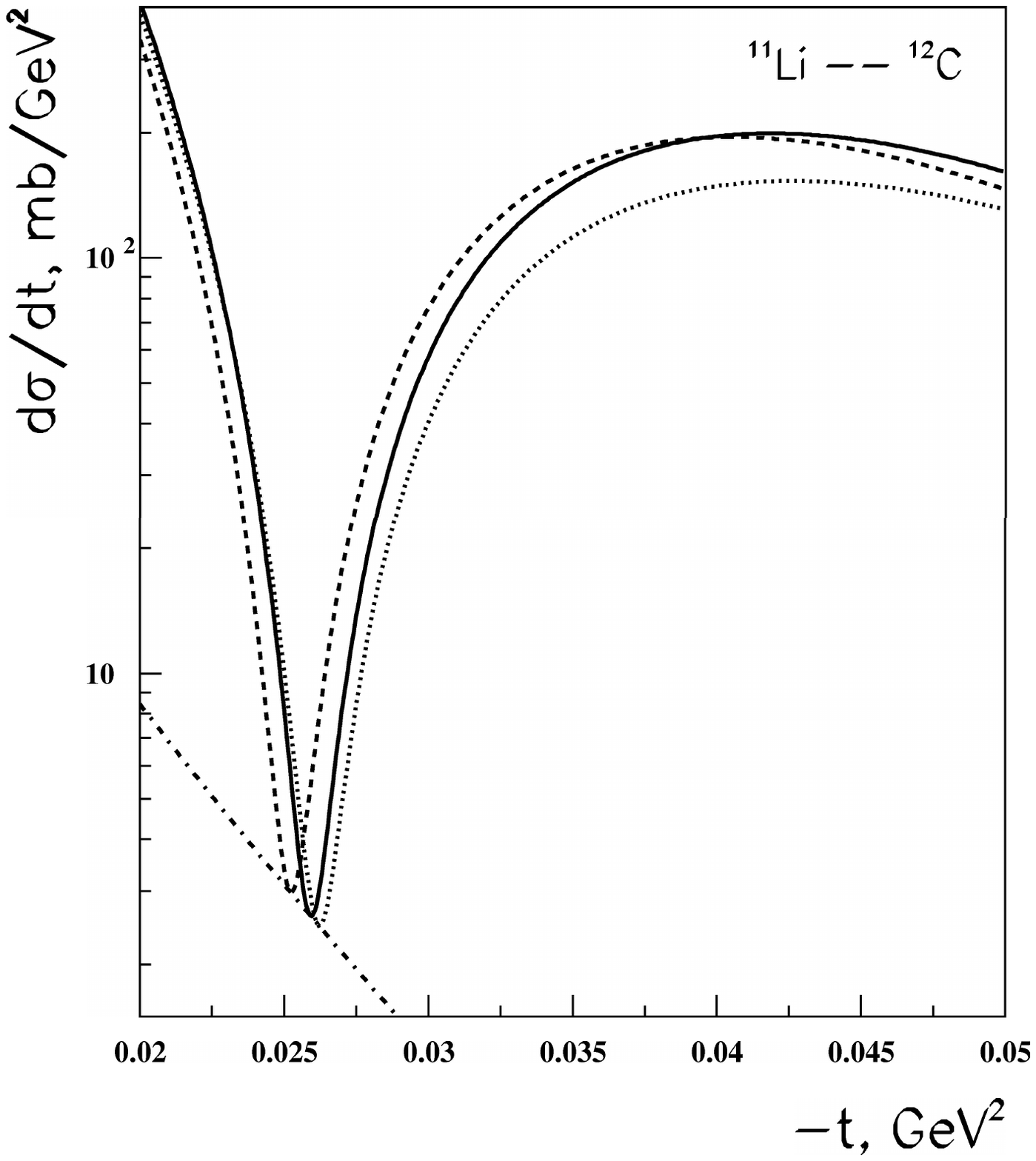}
\vskip -1.5cm
\caption{\footnotesize
The enlarged fragment of Fig.2. The notations are the same
as in Fig.2.
}
\end{figure}
The elastic differential cross section is evaluated through
the scattering amplitude $f$ as
$$
d\sigma\,=\,|\,f\,|^2 2\pi\,\sin\theta d\theta\,=\,
-\frac{\pi}{k^2}\,dt,~~~
dt=2 k^2\sin\theta d\theta,
$$
where $\theta$ and $k$ are the scattering angle
and the momentum conjugated to the relative distance
in the center of mass system,
$k = (M_A p_B-M_B p_A)/(M_A+M_B)$,
$t=(p-p^\prime)^2= 2 k^2(1-\cos\theta)$.
The amplitude $f$ is a sum of the Coulomb and
the short range strong ones, thus
$$
\frac{d\sigma}{dt}\,=\,\frac{\pi}{k^2}
\bigl|f_C(q)\,+\,F_{AB}^{el}(q)\bigr|^2\,=\,
\frac 1{4\pi}\,
\biggl|\,\frac{2\pi}{k}f_C(q)\,+\,i\phi(q)\,\biggr|^2,
$$
$$
\phi(q)\,=\,\int d^2b\,e^{iqb}[1-S_{AB}(b)].
$$
Given the fact that in the one photon exchange approximation
the Coulomb amplitude is real we have
\begin{equation}
\label{sum}
\frac{d\sigma}{dt}\,=\,
\frac 1{4\pi}\,
\biggl[\,\frac{4\pi^2}{k^2}f_C^2(q)\,+\,\phi^2(q)\,\biggr].
\end{equation}

\section{Numerical calculations}

The parameters of NN amplitude (\ref{f_NN}) have been chosen
to be $\sigma_{NN}^{tot}=43$~mb, $\beta = 0.2$~fm$^2$,
the NN total cross section being taken as an average
over $pp$ and $pn$ values.

We start from $^{12}$C -- $^{12}$C scattering. The distribution (\ref{HO}) has been
used with the value of the mean square radius $R_{\mathrm{rms}}=2.49$~fm fitted
from Monte-Carlo simulation of a $^{12}$C -- $^{12}$C collision in
Ref.~\cite{Merino:2009yj}. The results of the calculations are presented in Fig.1
The figure exhibits 2 maxima and 2 minima between them at the transferred momentum
interval $-t=0.001 - 0.15$GeV$^2$ (the point $-t=0$ is excluded for the singular
Coulomb contribution). As is seen from the figure the values the cross section
takes at the minima are due to the Coulomb interaction only. This is a consequence
of the fact that the real part of the strong NN amplitude is neglected.

The Fig.2 presents the results for
the scattering of the exotic halo nucleus $^{11}$Li on
$^{12}$C target obtained with the three
parametrizations~(\ref{params})
with $R_c=2.50$~fm, $R_v=5.04$~fm, $N_c=9$,
$N_v=2$~\cite{Hassan:2015dfa}.
Since the core part is common for all of them the Coulomb
contributions are identical and the difference should be
attributed to various shell structures supposed for
the neutrons in the halo.
To expose this difference
more explicitly the enlarged fragment of Fig.2 is presented in Fig.3.

\section{Conclusion}

We have carried out the calculations
of the differential cross sections
of $^{12}$C -- $^{12}$C and $^{11}$Li -- $^{12}$C scattering
in the complete Glauber theory with account for
the Coulomb contribution.
The Coulomb scattering plays an important role
in the nucleus-nucleus collisions especially
at the small transverse momenta.
In the case of $^{11}$Li -- $^{12}$C scattering we have
considered 3 parametrization for the halo
taken from Ref.~\cite{Hassan:2015dfa}. The results
show a notable difference between the obtained cross sections.


\begin{thebibliography}{**}
\bibitem{Glaub1}
R.J. Glauber, {\em Cross-sections in deuterium at high-energies}
Phys. Rev. \textbf{100}, 242 (1955).

\bibitem{Glauber:1970jm}
R.~J.~Glauber and G.~Matthiae,
{\em High-energy scattering of protons by nuclei},
Nucl. Phys. B \textbf{21}, 135 (1970).

\bibitem{Czyz:1969jg}
W.~Czyz and L.~C.~Maximon,
{\em High-energy, small angle elastic scattering
of strongly interacting composite particles},
Annals Phys. \textbf{52}, 59 (1969).

\bibitem{Alkhazov:2011ks}
G.~D.~Alkhazov and V.~V.~Sarantsev,
{\em Sensitivity of Cross Sections for Elastic Nucleus-Nucleus
Scattering to Halo Nucleus Density Distributions},
Phys. At. Nucl. \textbf{75}, 1544 (2012)
arXiv:1107.0533.

\bibitem{Alkhazov:2013mqa}
G.~D.~Alkhazov and V.~V.~Sarantsev,
{\em Sensitivity of reaction cross sections to halo
nucleus density distributions},
Phys. At. Nucl. \textbf{77}, 912-916 (2014)
arXiv:1312.0782.

\bibitem{Shabelski:2021iqk}
Y.~M.~Shabelski and A.~G.~Shuvaev,
{\em Generating function for nucleus-nucleus scattering
amplitudes in Glauber theory},
Phys. Rev. C \textbf{104}, no.6, 064607 (2021)
arXiv:2104.04943

\bibitem{Pajares:1983gw}
C.~Pajares and A.~V.~Ramallo,
{\em Effects of the multiple scattering structure
in the propagation of hadronic properties
in nucleus-nucleus collisions},
Phys. Rev. D \textbf{31}, 2800 (1985).

\bibitem{Braun:1988pk}
V.~M.~Braun and Y.~M.~Shabelski,
{\em Multiple Scattering Theory for Inelastic Processes},
Int. J. Mod. Phys. A \textbf{3}, 2417 (1988).

\bibitem{Alkhazov:2004syy}
G.~D.~Alkhazov, A.~V.~Dobrovolsky and A.~A.~Lobodenko,
{\em Matter density distributions and radii of light exotic nuclei
from intermediate-energy proton elastic scattering
and from interaction cross sections},
Nucl. Phys. A \textbf{734}, 361 (2004).

\bibitem{Ilieva:2012vhs}
S.~Ilieva, F.~Aksouh, G.~D.~Alkhazov, L.~Chulkov,
A.~V.~Dobrovolsky, P.~Egelhof, H.~Geissel, M.~Gorska,
A.~Inglessi, R.~Kanungo, \textit{et al.}
{\em Nuclear-matter density distribution in the neutron-rich
nuclei 12,14 Be from proton elastic scattering
in inverse kinematics},
Nucl. Phys. A \textbf{875}, 8 (2012).

\bibitem{Korsheninnikov:1997qta}
A.~A.~Korsheninnikov, E.~Y.~Nikolskii, C.~A.~Bertulani,
S.~Fukuda, T.~Kobayashi, E.~A.~Kuzmin, S.~Momota,
B.~G.~Novatskii, A.~A.~Ogloblin, A.~Ozawa, \textit{et al.}
{\em Scattering of radioactive nuclei 6 He and 3 H by protons:
Effects of neutron skin and halo in 6 He, 8 He, and 11 Li},
Nucl. Phys. A \textbf{617}, 45 (1997).

\bibitem{Zhukov:1993aw}
M.~V.~Zhukov, B.~V.~Danilin, D.~V.~Fedorov, J.~M.~Bang,
I.~J.~Thompson and J.~S.~Vaagen,
{\em Bound state properties of Borromean Halo nuclei:
He-6 and Li-11},
Phys. Rept. \textbf{231}, 151 (1993).

\bibitem{Hassan:2015dfa}
M.~A.~M.~Hassan, M.~S.~M.~Nour El-Din, A.~Ellithi,
E.~Ismail and H.~Hosny,
{\em The effect of halo nuclear density on reaction
cross-section for light ion collision},
Int. J. Mod. Phys. E \textbf{24}, 1550062 (2015).

\bibitem{Merino:2009yj}
C.~Merino, I.~S.~Novikov and Y.~M.~Shabelski,
{\em Nuclear Radii Calculations in Various Theoretical
Approaches for Nucleus-Nucleus Interactions},
Phys. Rev. C \textbf{80}, 064616 (2009).

\end{thebibliography}
\end{document}